\documentclass[twocolumn,prb,showpacs]{revtex4} % comment, if revtex is 
\usepackage{epsfig}
\usepackage{amssymb}

\begin{document}

\bibliographystyle{apsrev}

\title{Connection between vibrational dynamics and 
topological order in simple glasses}

\author{S.~I.~Simdyankin and M.~Dzugutov} 
\affiliation{Department of Numerical Analysis and Computer Science,
             Royal Institute of Technology, SE--100 44 Stockholm, Sweden}
 
\author{S.~N.~Taraskin and S.~R.~Elliott} 
\affiliation{Department of Chemistry, University of Cambridge,
             Lensfield Road, Cambridge CB2 1EW, UK}

\date{\today}

\begin{abstract}

We compare vibrational dynamics in two structurally distinct, simple
monatomic model glasses simulated by molecular dynamics: the
Lennard-Jones glass with an fcc-related structure and a glass with
predominantly icosahedral short-range order. The former, characterized
by a single local quasi-periodicity, supports only modes with acoustic
behaviour. In the latter, the presence of optic modes and two
incommensurate length scales is observed. This pattern of vibrational
dynamics is shown to be closely related to that of a Frank-Kasper
crystal having the same local topological order.
\end{abstract}


\maketitle

\section{Introduction}
\label{S1}
The nature of vibrational dynamics in topologically disordered solids
remains an unsolved problem of fundamental interest.
An important question is to what extent is the vibrational behaviour
determined by the atomic structure in such materials.
Liquids and glasses lack global symmetry, and, therefore, their
vibrational eigenstates are no longer phonons.
Nevertheless, long-wavelength phonon-like propagating acoustic plane
waves can be observed in these structures \cite{Taraskin_00:PRB_IR1,Taraskin_00:PRB_IR2}
up to a certain frequency, known as the Ioffe-Regel crossover
$\omega_{IR}$, beyond which they are heavily damped due to strong
scattering.
However, the spectral functions (e.g. the dynamical structure factor
$S(Q,\omega)$) are still peak-shaped even beyond $\omega_{IR}$ 
(see e.g. 
Ref.~\cite{Taraskin_00:PRB_IR1,Taraskin_00:PRB_IR2} 
and references therein).
Thus, the concept of dispersion, as peak-position frequency versus the
wavevector magnitude, is still valid in disordered structures
\cite{Taraskin_00:PRB_IR1,Taraskin_00:PRB_IR2,Grest_82,Grest_84,Taraskin_97:EPL,Sampoli_98,DelAnna_98}.
In glasses, it has been assumed that the position of the main maximum,
$Q_0$, of the static structure factor $S(Q)$ plays the role of the
first reciprocal-lattice point, with the first pseudo-Brillouin zone
extending up to $Q_0/2$ 
\cite{Grest_82,Grest_84,Hafner_87,Suck_92}.

However, the main maximum of $S(Q)$, manifesting the dominant
pseudo-periodicity in glasses, does not contain full information about
the glass structure.
Even in simple monatomic liquids and glasses, there exist distinct
types of short-range order (SRO) closely related to the respective
counterpart periodic structures \cite{Stillinger_85}.
Then, to what extent is the structural diversity of glasses reflected
in their vibrational properties?
In particular, can the distinctive features of the vibrational
dynamics characteristic of a crystalline structure be discerned in the
vibrational properties of a glass with similar local order?

A type of SRO that is most interesting in this respect is that based
on icosahedral coordination of the first shell of neighbors.
It is commonly associated with the glass-forming ability of simple
metallic systems \cite{Nelson_89}.
The frustration inherent in packing icosahedra in Euclidean (flat) 3D
space in the liquid state produces 6-fold defects which, on cooling,
form disclination lines.
The latter may align in space, forming globally ordered patterns,
e.g. Frank-Kasper periodic phases or morphologically related
quasicrystals.
Under sufficiently rapid cooling, the disclination lines get
entangled, forming a glass \cite{Nelson_89}.

The large numbers of atoms in the primitive cells of the Frank-Kasper
phases give rise to many optic vibrational modes.
Can these modes survive the lack of long-range order in the related
glass?
In addressing this issue, it is important to discriminate between the
optic modes arising as a result of a particular type of topological
SRO, of interest here, and those induced by the chemical SRO in a
system with more than one type of atom. The latter have been observed
in simple glassy alloys \cite{Hafner_83, Benmore_99}.

Here, we employ molecular dynamics to investigate the vibrational
dynamics in a monatomic model of a glass with icosahedral SRO (IC
glass) \cite{Dzugutov_92:glass}.
The thermodynamically stable solid phase on freezing for this system
is a dodecagonal quasicrystal \cite{Dzugutov_93:quasicrystal}, whereas
the energy-favored structure is the $\sigma$ phase
\cite{Dzugutov_97,Simdyankin_00,Roth_00}, a Frank-Kasper crystal
\cite{Frank_59}.

The rest of the paper is arranged as follows. 
In Sec.~\ref{S2}, we describe  simulation procedure. 
The results and their discussion are presented 
in Sec.~\ref{S3} and Sec.~\ref{S4}, respectively. 
The conclusions are given in Sec.~\ref{S5}. 

\section{Simulation}
\label{S2}

The IC glass was obtained in the course of a NVE molecular dynamics
(MD) simulation. A system of 16000 particles was cooled from the
liquid state above the melting point $T_m \simeq 0.5$ at a constant
density $\rho=0.8771$ (all quantities reported here are expressed in
the Lennard-Jones reduced units \cite{Allen_87} defining the pair
potential \cite{Dzugutov_92:glass}). 
This is the minimum energy density at zero pressure in the $\sigma$
phase \cite{Simdyankin_00}.
We cooled the system in a stepwise manner.
At each cooling step, the temperature of the system was reduced by scaling 
the velocities of the constituent particles.
Then the system was equilibrated at constant temperature until the
decrease of the potential energy was observed 
to cease,
indicating that a state of metastable equilibrium had been reached. 
Successful equilibration of this system in the metastable liquid
domain below the melting point was possible due to the complexity of
its crystallization pattern, significantly inhibiting the process
of crystalline nucleation. 
The equilibration time rapidly increased with decreasing temperature.
At $T \simeq 0.3$, this time exceeded the time of the simulation run
Below $T=0.3$, we found that the equilibration had not been completed
after the relaxation run of $t=10^7 \Delta t$, $\Delta t = 0.01$. 
At this point, the system, therefore, was assumed to be in the glassy
state.
Then it was cooled to the temperature $T=0.01$ at which the atoms
vibrate around their equilibrium positions and do not diffuse away,
and their vibrational dynamics was investigated.

\begin{figure}
\centerline{\epsfig{file=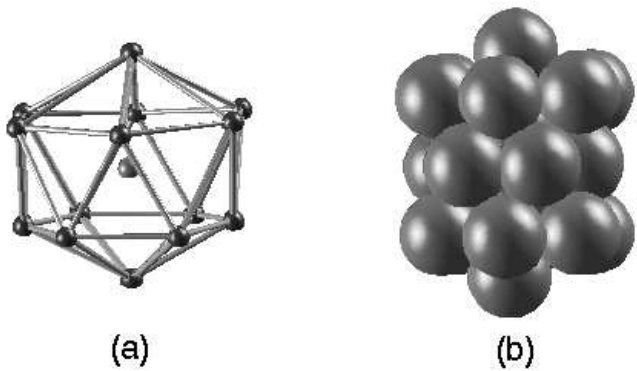,width=8.5cm,clip=}}
\caption{(a) Frank-Kasper $Z$14 polyhedron.  (b) A fragment of the
$-72^{\circ}$ disclination line (two interpenetrating $Z$14 elements)
as observed in the simulated IC glass.  }
\label{Z14}
\end{figure}

Mindful of the possible growth of crystalline grains in the process of
glass formation, we analyzed the evolution of the six-fold structural
elements, the Frank-Kasper $Z$14 polyhedra (shown in
Fig.~\ref{Z14}(a)).
In the process of Frank-Kasper crystallization, interpenetrating $Z$14
elements form $-72^{\circ}$ disclination lines (Fig.~\ref{Z14}(b))
which eventually thread the entire sample \cite{Nelson_89}.
Therefore, the extent of crystallization can be assessed by monitoring
the length of these disclination lines.
Table~\ref{StatZ14} demonstrates that the process of glass formation
in the IC system is not accompanied by any significant growth of the
crystalline phase.

\begin{table}[placement]
  \begin{tabular}{|r|r|r|r|}
     \hline 
number of $Z$14 atoms in & \multicolumn{3}{r|}{number of disclination lines} \\ 
\cline{2-4}
the disclination line & $T=1.0$ & $T=0.5$ & $T=0.3$ \\ \hline \hline
              1     &     149 &     260 &  599 \\ \hline
              2     &       3 &      27 &  145 \\ \hline
              3     &       0 &       4 &   28 \\ \hline 
              4     &       0 &       2 &    9 \\ \hline
              5     &       0 &       0 &    2 \\ \hline \hline
total number of Z14 atoms &155&     334 & 1019 \\ \hline
\end{tabular}
\caption{Temperature dependence of the statistics of $-72^{\circ}$
disclination lines in the IC glass.}
\label{StatZ14}
\end{table}

\begin{figure}
\centerline{\epsfig{file=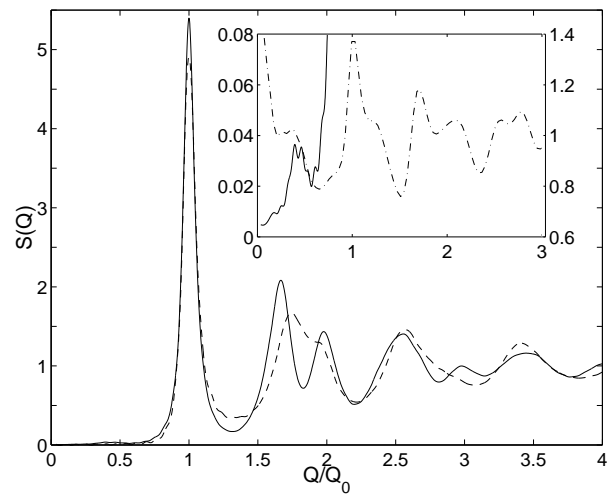,width=8cm,clip=}}
\caption{Static structure factors $S(Q)$ for the IC glass (solid line)
         and the LJ glass (dashed line).  The inset shows the
         small-$Q$ part of $S(Q)$ (solid line, left-hand scale) and
         the partial structure factor calculated for the centers of
         icosahedra (dash-dotted line, right-hand scale), both for the
         IC glass. ($Q_0 \rho^{-1/3} \simeq 7.17$ and $7.08$ for IC 
         and  LJ glass, respectively.)}
\label{SQs}
\end{figure}

As a reference, we simulated a monatomic Lennard-Jones (LJ) glass with
the SRO related to the face-centered cubic (fcc) structure. 
This glass was prepared at the density $\rho=1.0888$, corresponding to
the zero-pressure energy minimum of the fcc crystal with the LJ
potential truncated at $r_c=4$, by rapidly quenching a system of 16000
particles from a well-equilibrated liquid state at $T=3$ to $T=0.01$.
The structural distinction between the two glasses is demonstrated in
Fig.~\ref{SQs}.

In order to gain an insight into the vibrational properties of the IC
glass, we also studied the dynamics of a one-component model of the
$\sigma$ phase \cite{Simdyankin_00}.
The latter model utilized the same pair potential as the IC-glass
model.
Both models were studied at the same density.

\begin{figure*}
\centerline{\epsfig{file=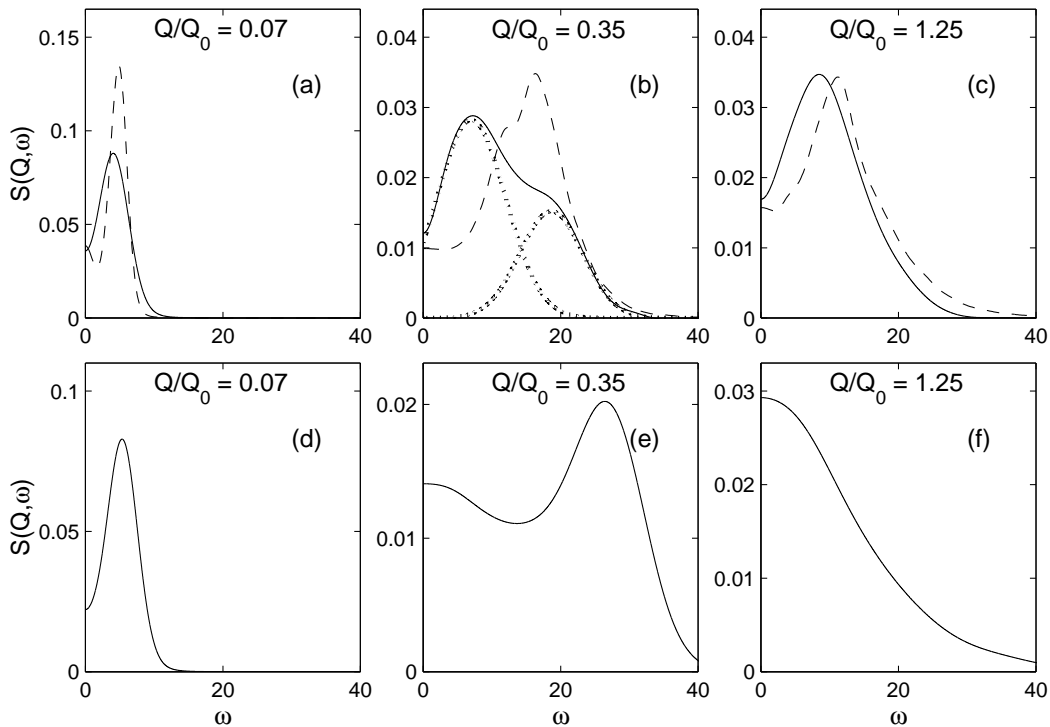,width=14cm,clip=}}
\caption{Dynamical structure factors in the simulated systems at three
values of wavenumber as marked. (a)-(c): IC glass, solid line; dashed
line, $\sigma$ phase.  In (b), the dotted curves show two Gaussians,
the sum of which is fitted to the solid curve; this fit, if plotted,
is indistinguishable from the solid curve. (d)-(f): LJ glass.  }
\label{SQos}
\end{figure*}

We associated longitudinal vibrational modes with non-zero frequency
maxima in the dynamical structure factor, $S(Q,\omega)$, the spectrum of
the density-fluctuation time-autocorrelation function
\cite{Hansen_86}:
%\begin{equation}
$
S(Q,\omega) = N^{-1}
\int_0^\infty dt \; e^{i\omega t} \langle \rho({\bf Q},t)
\rho(-{\bf Q},0) \rangle,
%\label{S(Q,omega)}
$
%\end{equation}
where 
%\begin{equation}
$
\rho({\bf Q},t)=\sum_{k=1}^{N}\exp[-i{\bf Q}\cdot{\bf r}_k(t)]
$
%\end{equation}
and ${\bf r}_k(t)$ is the positional vector of the particle $k$.
For transverse vibrational modes, we calculated the spectra of the
transverse current fluctuations \cite{Hansen_86}:
%\begin{equation}
$
C_t(Q,\omega) = Q^2N^{-1} \int_0^\infty dt \; e^{i\omega t} 
\langle j_t({\bf Q},t) j_t(-{\bf Q},0) \rangle,
%\label{C(Q,omega)}
$
%\end{equation}
where $j_t({\bf Q},t)$ is the transverse component of the
Fourier-transform of the local current:
%\begin{equation}
$
j_t({\bf Q},t) = 
\sum_{k=1}^{N} ({\bf e}_t\cdot{\bf v}_k(t))
\exp[-i{\bf Q} \cdot {\bf r}_k(t)],
%\label{j(Q,t)}
$
%\end{equation}
${\bf e}_t \perp {\bf Q}$ and ${\bf v}_k(t)$ is the velocity of
particle $k$.
Finite-time truncation effects in the spectra of the time-correlation
functions were reduced by convolution with a Gaussian window function.

\begin{figure*}
\centerline{\epsfig{file=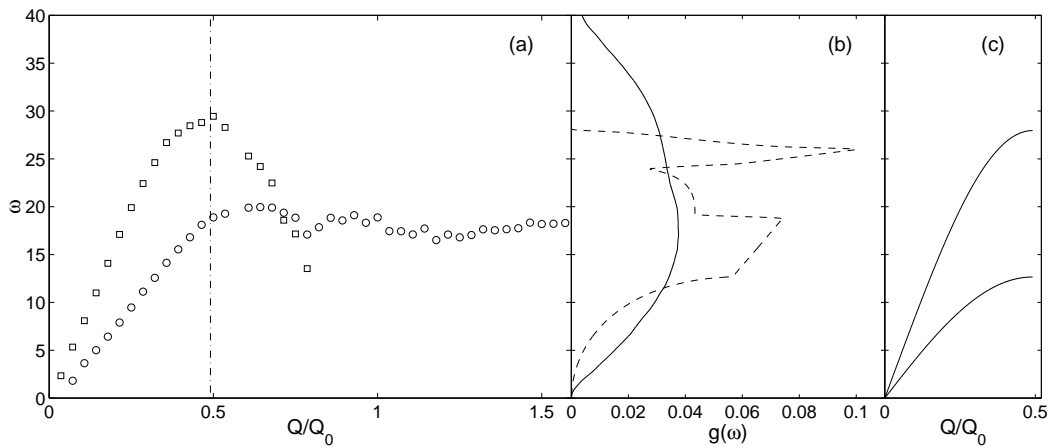,width=14cm,clip=}}
\caption{(a) Dispersion of vibrational excitations for the LJ
glass. Open squares: longitudinal branch.  Open circles: transverse
branch.  Dash-dotted line: the first Brillouin-zone boundary for the
fcc structure in the $(111)$ symmetry direction.  (b) Vibrational
density of states for the LJ glass (solid curve) and for the fcc
structure (dashed curve).  (c) Phonon-dispersion curves for the fcc
structure in the $(111)$ symmetry direction.  }
\label{dispLJ}
\end{figure*}

\begin{figure*}
\centerline{\epsfig{file=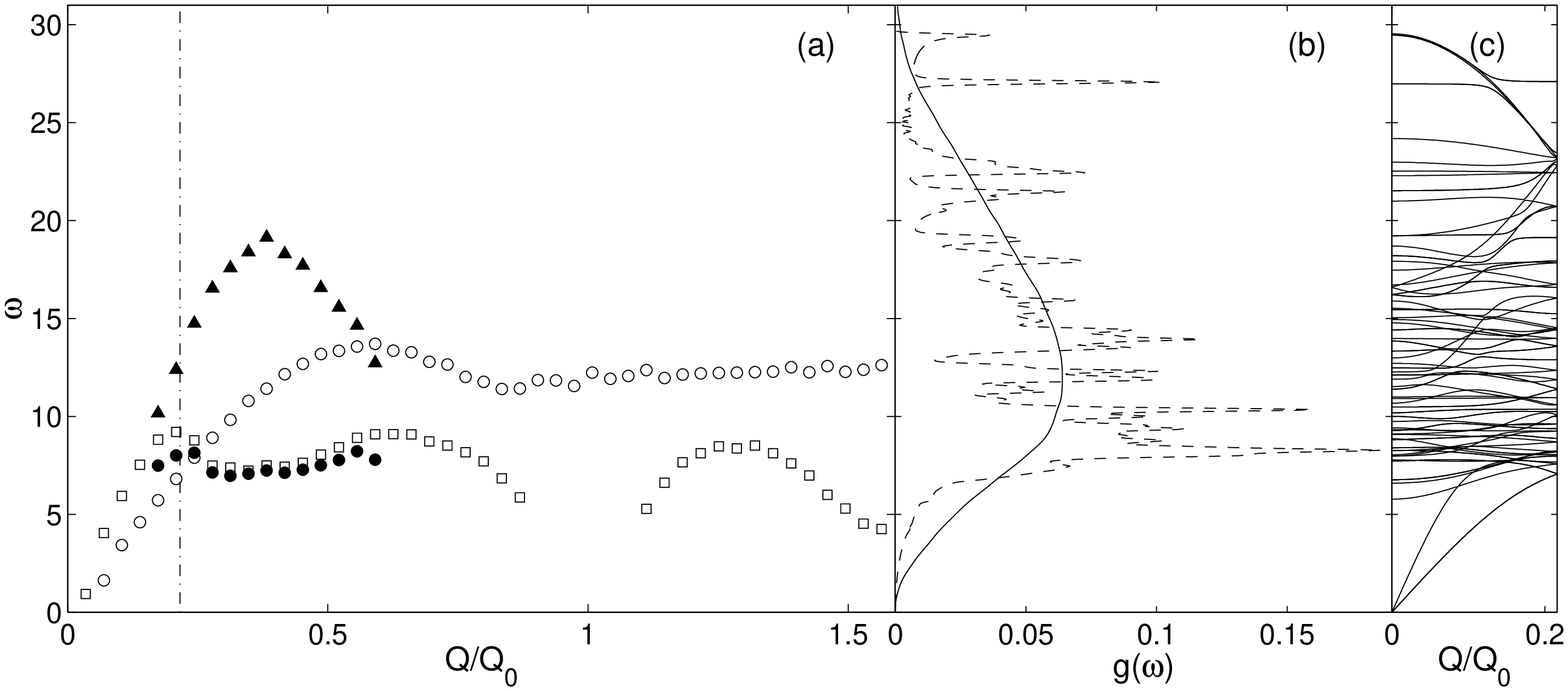,width=14cm,clip=}}
\caption{(a) Dispersion of vibrational excitations 
         in the IC glass. 
         Open squares: longitudinal branch.  Open circles: transverse
         branch.  Solid circles and triangles: maxima positions of the
         two Gaussians approximating $S(Q,\omega)$.  Dash-dotted
         line: the first Brillouin zone boundary for the
         $\sigma$-phase structure in the $(001)$ symmetry direction.
         (b) Vibrational densities of states of the IC glass (solid
         line), and the $\sigma$-phase structure (dashed line).  (c)
         Phonon-dispersion curves for the $\sigma$-phase structure in
         the $(001)$ symmetry direction. }
\label{dispIC}
\end{figure*}

\section{Results}
\label{S3}
Fig.~\ref{SQos} shows $S(Q,\omega)$ for the two glassy models, as well
as the $\sigma$ phase, for several representative values of $Q$.
For sufficiently small $Q$, both glassy systems clearly exhibit acoustic
modes which manifest themselves as pronounced  Brillouin peaks in
$S(Q,\omega)$, Fig.~\ref{SQos}(a).
A marked difference between the two sets of curves arises for
$Q\gtrsim0.2\,Q_0$.
The width of the Brillouin peak in the LJ glass becomes comparable
with that of the whole vibrational spectrum (see Fig.~\ref{SQos}(e))
and for larger $Q$ the peak disappears (Fig.~\ref{SQos}(f)).
For the IC glass, a well-defined Brillouin peak persists well beyond
$Q_0$ (Fig.~\ref{SQos}(c)).
Another important feature of the IC glass is that, for $Q>0.2\,Q_0$,
the Brillouin maximum develops a well-defined shoulder on the
high-frequency side (Fig.~\ref{SQos}(b)).
We analyzed this structure of the Brillouin peak by fitting it with
the sum of two Gaussians.

The assumed interpretation of the shoulder observed in the Brillouin
peak in $S(Q,\omega)$ of the IC glass as indicating the presence of a
separate maximum can be corroborated by a comparison with the
structure of $S(Q,\omega)$ calculated for the $\sigma$ phase which is
also shown in Fig.~\ref{SQos}(a)-(c).
$S(Q,\omega)$ for the $\sigma$ phase was computed at $T=0.8$ in the
same way as for the glassy systems: the averaging was
performed over all wavevectors ${\bf Q}$ confined to a narrow
spherical shell in  reciprocal space, such that $|{\bf Q}| \simeq
Q$.
For $Q/Q_0=0.35$, \(S(Q,\omega)\) exhibits a pronounced peak at the
frequency close to the position of the shoulder observed
in $S(Q,\omega)$ of the glassy phase.
The two sets of curves in Fig~\ref{SQos}(a)-(c) clearly show certain
similarities: the number of peaks is the same and their corresponding
positions are close.
This leads us to the conclusion that the shape of the dynamical
structure factor in the frequency domain for the IC glass is
generically related to that for its crystalline counterpart, the
$\sigma$ phase.
Namely, the peaks in \( S(Q,\omega) \) for the IC glass originate from
the corresponding peaks for the $\sigma$ phase, which are broadened
and shifted by disorder-induced level-repelling effects
\cite{Taraskin_00:PRL}.
The level-repelling and hybridization effects describe a
reconstruction of the crystalline vibrational spectrum to give the
disordered one.
These should give rise to a shift of the low-frequency crystalline
peaks downwards in frequency but upwards for the high-frequency peaks
(see Ref.~\cite{Taraskin_00:PRL} for more detail).
Such tendencies are indeed seen in Fig.~\ref{SQos}(a)-(c) for the
relative positions of the peaks for the $\sigma$ phase and the IC
glass.
Therefore, our results support the conjecture that disorder-induced
level-repelling effects in vibrational states in crystalline
counterparts of topologically disordered structures, such as the IC
glass, are responsible for the main features of their vibrational
spectrum.

The dispersion relations as derived from the
positions of the respective Brillouin maxima for the LJ and IC models
are shown in Figs.~\ref{dispLJ} and \ref{dispIC}, respectively.
The patterns corresponding to the longitudinal modes in the two glassy
systems are strikingly different.
The longitudinal dispersion curve for the LJ glass demonstrates a pattern
typical for acoustic modes, with a maximum centered at around $Q_0/2$,
Fig.~\ref{dispLJ}(a).
The dispersion pattern of the single Brillouin peak observed in  
$S(Q,\omega)$ for the IC system for $Q<0.2\,Q_0$ can also be 
identified with acoustic behavior. 
For larger values of $Q$, we identify two separate dispersion branches
associated with the positions of the maxima of the two Gaussians 
approximating $S(Q,\omega)$, as shown in Fig.~\ref{dispIC}(a).
The main subpeak of the Brillouin maxima exhibits a very weak
$Q$-dependence, resembling the behavior of the optic modes in
crystalline solids (Fig.~\ref{dispIC}).
The shoulder observed in the Brillouin peaks of the IC system 
gives rise to a higher-frequency dispersion branch which exhibits 
a maximum.
However, in contrast to the acoustic-type dispersion observed in the LJ 
glass, the maximum is located at around $0.4\,Q_0$.

\section{Discussion}
\label{S4}
The pronounced differences in the behavior of the longitudinal modes
in these two model glasses can be rationalized by considering the
respective modes in the crystalline counterparts - the $\sigma$ phase
in the case of the IC glass, and the fcc crystal in the case of the LJ
glass.

The LJ glass is characterized by a single dominant structural repeat
distance, manifested by the  first peak of $S(Q)$. It corresponds
to the interplanar spacing along the $(111)$ symmetry direction in the
fcc structure (i.e., the apex-basal-plane distance in a close-packed
tetrahedron). $Q_0$ being the first reciprocal-lattice point, the
first Brillouin-zone boundary is at $Q_0/2$, which corresponds to the
maximum in the dispersion curve of the longitudinal acoustic (LA)
modes  (Fig.~\ref{dispLJ}(a)).  As a result of this structural
simplicity, the vibrational dispersion pattern in the LJ glass is also
simple, with only LA and transverse branches, as in the counterpart
crystalline (fcc) structure (Fig.~\ref{dispLJ}(c)) that has just one
atom as the basis in the primitive unit cell.

The pattern of longitudinal vibrational modes in the IC glass is
considerably more complex (Fig.~\ref{dispIC}(a)), displaying a
low-frequency weakly dispersive branch with the first maximum at
$Q/Q_0 \simeq 0.2$.
Its dispersion behavior clearly resembles that of the optic 
modes which dominate this range of frequencies in the vibrational 
spectra of the \(\sigma\) phase (Fig.~\ref{dispIC}(c)). 
In the same domain of $Q$, the IC glass also exhibits a
higher-frequency branch associated with the high-frequency shoulder of
the Brillouin peak.
As has been mentioned earlier, this shoulder mirrors a peak in
$S(Q,\omega)$ of the high-temperature $\sigma$ phase
(Fig.~\ref{SQos}(b)).
The latter  manifests  optic modes and therefore we
conclude that  optic-type modes dominate the vibrational dynamics
in the IC glass for $Q/Q_0>0.2$.
The position of the maximum in the high-frequency branch, $Q/Q_0
\simeq 0.4$, is twice that in the low-frequency branch, $Q/Q_0 \simeq
0.2$. 
However, both branches are suppressed at $Q=Q_0$ by the influence of
the main peak of $S(Q)$.

The appearance of modes in the IC glass, mirroring the optic modes in
the corresponding $\sigma$-phase crystal (Fig.~\ref{dispIC}(c)), is a
consequence of the type of topological SRO that exists in both
structures.
The first Brillouin-zone boundary in the $\sigma$ phase in the $(001)$
symmetry direction is at $Q\simeq0.2\,Q_0$ (Fig.~\ref{dispIC}(c)),
corresponding very closely to the first pseudo-Brillouin-zone boundary
value found in the IC glass (Fig.~\ref{dispIC}(a)).
The behavior observed in the IC glass, that the low-frequency branch
first peaks at $Q\simeq 0.2\,Q_0$ whereas the high-frequency branch
peaks at $Q\simeq 0.4\, Q_0$, is typical for nonsymmorphic crystalline
structures, of which the $\sigma$ phase is an example (space group
$P4_2/mnm$), for ${\bf Q}$ in the direction of a screw axis
\cite{Steurer_93}.
Some modes sense any particular (pseudo-) periodic atomic-density
fluctuations, whereas some modes sense alternate such atomic-layer
spacings and therefore the apparent first Brillouin-zone boundaries
differ by a factor of two.
On the other hand, the dominant length scale in the IC glass that is
manifested by the main peak of $S(Q)$ corresponds to the layer
periodicity in the (111) direction of the $\sigma$ phase (the same
apex-basal-plane distance in tetrahedra as in fcc).

Optic modes in the IC glass give no contribution to $S(Q,\omega)$ for
$Q/Q_0<0.2$.  
This is not surprising because the dynamical structure factor
describes the response of the system to plane wave excitations.  Even
in crystals, the optic phonons are not simple plane waves (the atoms
inside the unit cell are not displaced according to a plane wave), and
in the long-wavelength limit ($Q \to 0$) only acoustic modes respond
to plane-wave excitations
\cite{Taraskin_00:PRB_IR1,Taraskin_00:PRB_IR2}.
The signature of optic-like excitations appears in the dynamic
structure factor of the IC glass only when the wavelength becomes
comparable with the size of structural units responsible for this
effect.

The dispersion curves for the IC glass thus indicate the existence of
an additional pseudo-periodic interlayer spacing incommensurate with
that manifested by the main peak of $S(Q)$. The latter corresponds to
the same pseudo-period in the atomic-density fluctuations as for the
LJ glass.
The second length scale is related to the interlayer spacing in aggregates
of interpenetrating icosahedral units, as along the $(001)$-axis in the
$\sigma$ phase.
This assertion is confirmed by the fact that both the total $S(Q)$ and
the partial structure factor calculated only for the centers of
icosahedra exhibit a prepeak at $Q\simeq 0.4\,Q_0$ (see the inset in
Fig.~\ref{SQs}).
The presence of an additional length scale in the IC glass appears to
agree with recent observations of medium-range order in
paracrystalline amorphous models \cite{Treacy_98,Treacy_00}.
This length scale is associated with a particular type of structural
unit which is morphologically close to those found in the
counterpart crystalline structure.
In a reciprocal-space picture, the effect of these structural units
can be analogous to that of the ``quasi-Bragg planes''
\cite{Uhlherr_94,Gaskell_96} used for explanation of the first sharp
diffraction peak in $S(Q)$ of amorphous solids.

The difference in the behaviour of collective vibrational modes in the
two glasses is also reflected in their densities of states,
Figs.~\ref{dispLJ}(b),~\ref{dispIC}(b).
On the other hand, the curves for the vibrational density of states in
both glasses appear to be consistent with the respective curves
calculated for the crystalline counterparts.

The dispersion patterns of the transverse modes are qualitatively
similar (Figs.~\ref{dispLJ}(a),~\ref{dispIC}(a)).
This implies that either these modes are insensitive to the structure
or, more likely, any structure effects cannot be shown using
$C_t(Q,\omega)$. This will be analyzed elsewhere.

\section{Conclusion}
\label{S5}
In conclusion, we have demonstrated that structural distinctions in
simple monatomic glasses give rise to an appreciable difference in the
behaviour of vibrational excitations.
We have shown that the main features 
in the dynamical structure factor of the IC glass 
are generically related to those of its 
crystalline counterpart, the \(\sigma\) phase.  
The other distinctive features of vibrational dynamics displayed by
the IC glass, namely, the presence of optic-like modes and an
additional pseudo-Brillouin zone boundary incommensurate with $Q_0$,
can be accounted for by the specific structure of aggregates of
interpenetrating icosahedra characteristic of the $\sigma$ phase,
which was previously shown \cite{Simdyankin_00} to be a good
crystalline counterpart of the IC glass.
This observation implies that the structural similarity between a
glass and its crystalline counterpart extends well beyond the shell of
nearest neighbors.
The existence of an additional incommensurate length scale and optic
modes may be a generic feature of glasses having complex counterpart
crystalline structures.

\section*{Acknowledgments}

S.I.S. and M.D. thank Trinity College, Cambridge for hospitality, and
acknowledge support from the following Swedish research funds:
Natural Science Research Foundation (NFR), 
Technical Research Foundation (TFR), and 
Network for Applied Mathematics (NTM).
S.N.T. is grateful to EPSRC for support.

%\bibliography{archive}

\begin{thebibliography}{10}
\expandafter\ifx\csname bibnamefont\endcsname\relax
  \def\bibnamefont#1{#1}\fi
\expandafter\ifx\csname bibfnamefont\endcsname\relax
  \def\bibfnamefont#1{#1}\fi
\expandafter\ifx\csname url\endcsname\relax
  \def\url#1{\texttt{#1}}\fi
\expandafter\ifx\csname urlprefix\endcsname\relax\def\urlprefix{URL }\fi
\expandafter\ifx\csname bibinfo\endcsname\relax \def\bibinfo#1#2{#2}\fi
\expandafter\ifx\csname eprint\endcsname\relax \def\eprint#1{#1}\fi

\bibitem{Taraskin_00:PRB_IR1}
\bibinfo{author}{\bibfnamefont{S.~N.} \bibnamefont{Taraskin}} \bibnamefont{and}
  \bibinfo{author}{\bibfnamefont{S.~R.} \bibnamefont{Elliott}},
  \bibinfo{journal}{Phys. Rev. B} \textbf{\bibinfo{volume}{61}},
  \bibinfo{pages}{12017} (\bibinfo{year}{2000}).

\bibitem{Taraskin_00:PRB_IR2}
\bibinfo{author}{\bibfnamefont{S.~N.} \bibnamefont{Taraskin}} \bibnamefont{and}
  \bibinfo{author}{\bibfnamefont{S.~R.} \bibnamefont{Elliott}},
  \bibinfo{journal}{Phys. Rev. B} \textbf{\bibinfo{volume}{61}},
  \bibinfo{pages}{12031} (\bibinfo{year}{2000}).

\bibitem{Grest_82}
\bibinfo{author}{\bibfnamefont{G.~S.} \bibnamefont{Grest}},
  \bibinfo{author}{\bibfnamefont{S.~R.} \bibnamefont{Nagel}}, \bibnamefont{and}
  \bibinfo{author}{\bibfnamefont{A.}~\bibnamefont{Rahman}},
  \bibinfo{journal}{Phys. Rev. Lett.} \textbf{\bibinfo{volume}{49}},
  \bibinfo{pages}{1271} (\bibinfo{year}{1982}).

\bibitem{Grest_84}
\bibinfo{author}{\bibfnamefont{G.~S.} \bibnamefont{Grest}},
  \bibinfo{author}{\bibfnamefont{S.~R.} \bibnamefont{Nagel}}, \bibnamefont{and}
  \bibinfo{author}{\bibfnamefont{A.}~\bibnamefont{Rahman}},
  \bibinfo{journal}{Phys. Rev. B} \textbf{\bibinfo{volume}{29}},
  \bibinfo{pages}{5968} (\bibinfo{year}{1984}).

\bibitem{Taraskin_97:EPL}
\bibinfo{author}{\bibfnamefont{S.~N.} \bibnamefont{Taraskin}} \bibnamefont{and}
  \bibinfo{author}{\bibfnamefont{S.~R.} \bibnamefont{Elliott}},
  \bibinfo{journal}{Europhys. Lett.} \textbf{\bibinfo{volume}{39}},
  \bibinfo{pages}{37} (\bibinfo{year}{1997}).

\bibitem{Sampoli_98}
\bibinfo{author}{\bibfnamefont{M.}~\bibnamefont{Sampoli}},
  \bibinfo{author}{\bibfnamefont{P.}~\bibnamefont{Benassi}},
  \bibinfo{author}{\bibfnamefont{R.}~\bibnamefont{Dell'Anna}},
  \bibinfo{author}{\bibfnamefont{V.}~\bibnamefont{Mazzacurati}},
  \bibnamefont{and} \bibinfo{author}{\bibfnamefont{G.}~\bibnamefont{Ruocco}},
  \bibinfo{journal}{Phil. Mag. B} \textbf{\bibinfo{volume}{77}},
  \bibinfo{pages}{473} (\bibinfo{year}{1998}).

\bibitem{DelAnna_98}
\bibinfo{author}{\bibfnamefont{R.}~\bibnamefont{Dell'Anna}},
  \bibinfo{author}{\bibfnamefont{G.}~\bibnamefont{Ruocco}},
  \bibinfo{author}{\bibfnamefont{M.}~\bibnamefont{Sampoli}}, \bibnamefont{and}
  \bibinfo{author}{\bibfnamefont{G.}~\bibnamefont{Viliani}},
  \bibinfo{journal}{Phys. Rev. Lett.} \textbf{\bibinfo{volume}{80}},
  \bibinfo{pages}{1238} (\bibinfo{year}{1998}).

\bibitem{Hafner_87}
\bibinfo{author}{\bibfnamefont{J.}~\bibnamefont{Hafner}},
  \emph{\bibinfo{title}{From Hamiltonians to Phase Diagrams}}
  (\bibinfo{publisher}{Springer-Verlag}, \bibinfo{address}{Berlin},
  \bibinfo{year}{1987}).

\bibitem{Suck_92}
\bibinfo{author}{\bibfnamefont{J.-B.} \bibnamefont{Suck}},
  \bibinfo{author}{\bibfnamefont{P.~A.} \bibnamefont{Egelstaff}},
  \bibinfo{author}{\bibfnamefont{R.~A.} \bibnamefont{Robinson}},
  \bibinfo{author}{\bibfnamefont{D.~S.} \bibnamefont{Silvia}},
  \bibnamefont{and} \bibinfo{author}{\bibfnamefont{A.~D.}
  \bibnamefont{Taylor}}, \bibinfo{journal}{J. Non-Cryst. Solids}
  \textbf{\bibinfo{volume}{150}}, \bibinfo{pages}{245} (\bibinfo{year}{1992}).

\bibitem{Stillinger_85}
\bibinfo{author}{\bibfnamefont{F.~H.} \bibnamefont{Stillinger}}
  \bibnamefont{and} \bibinfo{author}{\bibfnamefont{R.~A.}
  \bibnamefont{LaViolette}}, \bibinfo{journal}{J. Chem. Phys.}
  \textbf{\bibinfo{volume}{83}}, \bibinfo{pages}{6413} (\bibinfo{year}{1985}).

\bibitem{Nelson_89}
\bibinfo{author}{\bibfnamefont{D.~R.} \bibnamefont{Nelson}} \bibnamefont{and}
  \bibinfo{author}{\bibfnamefont{F.}~\bibnamefont{Spaepen}},
  \bibinfo{journal}{Solid State Phys.} \textbf{\bibinfo{volume}{42}},
  \bibinfo{pages}{1} (\bibinfo{year}{1989}).

\bibitem{Hafner_83}
\bibinfo{author}{\bibfnamefont{J.}~\bibnamefont{Hafner}}, \bibinfo{journal}{J.
  Phys. C} \textbf{\bibinfo{volume}{16}}, \bibinfo{pages}{5773}
  (\bibinfo{year}{1983}).

\bibitem{Benmore_99}
\bibinfo{author}{\bibfnamefont{C.~J.} \bibnamefont{Benmore}},
  \bibinfo{author}{\bibfnamefont{S.}~\bibnamefont{Sweeney}},
  \bibinfo{author}{\bibfnamefont{R.~A.} \bibnamefont{Robinson}},
  \bibinfo{author}{\bibfnamefont{P.~A.} \bibnamefont{Egelstaff}},
  \bibnamefont{and} \bibinfo{author}{\bibfnamefont{J.-B.} \bibnamefont{Suck}},
  \bibinfo{journal}{J. Phys.: Condens. Matter} \textbf{\bibinfo{volume}{11}},
  \bibinfo{pages}{7079} (\bibinfo{year}{1999}).

\bibitem{Dzugutov_92:glass}
\bibinfo{author}{\bibfnamefont{M.}~\bibnamefont{Dzugutov}},
  \bibinfo{journal}{Phys. Rev. A} \textbf{\bibinfo{volume}{46}},
  \bibinfo{pages}{R2984} (\bibinfo{year}{1992}).

\bibitem{Dzugutov_93:quasicrystal}
\bibinfo{author}{\bibfnamefont{M.}~\bibnamefont{Dzugutov}},
  \bibinfo{journal}{Phys. Rev. Lett.} \textbf{\bibinfo{volume}{70}},
  \bibinfo{pages}{2924} (\bibinfo{year}{1993}).

\bibitem{Dzugutov_97}
\bibinfo{author}{\bibfnamefont{M.}~\bibnamefont{Dzugutov}},
  \bibinfo{journal}{Phys. Rev. Lett.} \textbf{\bibinfo{volume}{79}},
  \bibinfo{pages}{4043} (\bibinfo{year}{1997}).

\bibitem{Simdyankin_00}
\bibinfo{author}{\bibfnamefont{S.~I.} \bibnamefont{Simdyankin}},
  \bibinfo{author}{\bibfnamefont{S.~N.} \bibnamefont{Taraskin}},
  \bibinfo{author}{\bibfnamefont{M.}~\bibnamefont{Dzugutov}}, \bibnamefont{and}
  \bibinfo{author}{\bibfnamefont{S.~R.} \bibnamefont{Elliott}},
  \bibinfo{journal}{Phys. Rev. B} \textbf{\bibinfo{volume}{62}},
  \bibinfo{pages}{3223} (\bibinfo{year}{2000}).

\bibitem{Roth_00}
\bibinfo{author}{\bibfnamefont{J.}~\bibnamefont{Roth}} \bibnamefont{and}
  \bibinfo{author}{\bibfnamefont{A.~R.} \bibnamefont{Denton}},
  \bibinfo{journal}{Phys. Rev. E} \textbf{\bibinfo{volume}{61}},
  \bibinfo{pages}{6845} (\bibinfo{year}{2000}).

\bibitem{Frank_59}
\bibinfo{author}{\bibfnamefont{F.~C.} \bibnamefont{Frank}} \bibnamefont{and}
  \bibinfo{author}{\bibfnamefont{J.~S.} \bibnamefont{Kasper}},
  \bibinfo{journal}{Acta Cryst.} \textbf{\bibinfo{volume}{12}},
  \bibinfo{pages}{483} (\bibinfo{year}{1959}).

\bibitem{Allen_87}
\bibinfo{author}{\bibfnamefont{M.~P.} \bibnamefont{Allen}} \bibnamefont{and}
  \bibinfo{author}{\bibfnamefont{D.~J.} \bibnamefont{Tildesley}},
  \emph{\bibinfo{title}{Computer Simulation of Liquids}}
  (\bibinfo{publisher}{Clarendon Press}, \bibinfo{address}{Oxford},
  \bibinfo{year}{1987}).

\bibitem{Hansen_86}
\bibinfo{author}{\bibfnamefont{J.-P.} \bibnamefont{Hansen}} \bibnamefont{and}
  \bibinfo{author}{\bibfnamefont{I.~R.} \bibnamefont{McDonald}},
  \emph{\bibinfo{title}{Theory of Simple Liquids}}
  (\bibinfo{publisher}{Academic Press}, \bibinfo{address}{London},
  \bibinfo{year}{1986}), 2nd ed.

\bibitem{Taraskin_00:PRL}
\bibinfo{author}{\bibfnamefont{S.~N.} \bibnamefont{Taraskin}},
  \bibinfo{author}{\bibfnamefont{Y.~L.} \bibnamefont{Loh}},
  \bibinfo{author}{\bibfnamefont{G.}~\bibnamefont{Natarajan}},
  \bibnamefont{and} \bibinfo{author}{\bibfnamefont{S.~R.}
  \bibnamefont{Elliott}}, \bibinfo{journal}{Phys. Rev. Lett.}
  \bibinfo{note}{Submitted}.

\bibitem{Steurer_93}
\bibinfo{author}{\bibfnamefont{W.}~\bibnamefont{Steurer}}, in
  \emph{\bibinfo{booktitle}{Structure of Solids}}, edited by
  \bibinfo{editor}{\bibfnamefont{V.}~\bibnamefont{Gerod}}
  (\bibinfo{publisher}{VCH}, \bibinfo{address}{Weinheim},
  \bibinfo{year}{1993}).

\bibitem{Treacy_98}
\bibinfo{author}{\bibfnamefont{M.~M.~J.} \bibnamefont{Treacy}},
  \bibinfo{author}{\bibfnamefont{J.~M.} \bibnamefont{Gibson}},
  \bibnamefont{and} \bibinfo{author}{\bibfnamefont{P.~J.}
  \bibnamefont{Keblinski}}, \bibinfo{journal}{J. Non-Cryst. Solids}
  \textbf{\bibinfo{volume}{231}}, \bibinfo{pages}{99} (\bibinfo{year}{1998}).

\bibitem{Treacy_00}
\bibinfo{author}{\bibfnamefont{M.~M.~J.} \bibnamefont{Treacy}},
  \bibinfo{author}{\bibfnamefont{P.~M.} \bibnamefont{Voyles}},
  \bibnamefont{and} \bibinfo{author}{\bibfnamefont{J.~M.}
  \bibnamefont{Gibson}}, \bibinfo{journal}{J. Non-Cryst. Solids}
  \textbf{\bibinfo{volume}{266-269}}, \bibinfo{pages}{150}
  (\bibinfo{year}{2000}).

\bibitem{Uhlherr_94}
\bibinfo{author}{\bibfnamefont{A.}~\bibnamefont{Uhlherr}} \bibnamefont{and}
  \bibinfo{author}{\bibfnamefont{S.~R.} \bibnamefont{Elliott}},
  \bibinfo{journal}{J. Phys.: Cond. Mat.} \textbf{\bibinfo{volume}{6}},
  \bibinfo{pages}{L99} (\bibinfo{year}{1994}).

\bibitem{Gaskell_96}
\bibinfo{author}{\bibfnamefont{P.~H.} \bibnamefont{Gaskell}} \bibnamefont{and}
  \bibinfo{author}{\bibfnamefont{D.~J.} \bibnamefont{Wallis}},
  \bibinfo{journal}{Phys. Rev. Lett.}
  \textbf{\bibinfo{volume}{76}}, \bibinfo{pages}{66}
  (\bibinfo{year}{1996}).

\end{thebibliography}

\end{document}